\documentclass[pra,
                twocolumn,
                preprintnumbers,
                amsmath,
                amssymb,
                showpacs,
                superscriptaddress]{revtex4-1}
\usepackage{graphicx} % Include figure files
\usepackage{dcolumn}  % Align table columns on decimal point
\usepackage{bm}       % bold math
\usepackage{amsfonts}
\usepackage{mathtools}
\usepackage[colorlinks=true,linkcolor=blue,citecolor=red,urlcolor=blue]{hyperref}
\usepackage{color}
\usepackage{natbib}
\newcommand{\Eion}{\mbox{$E_{\mbox{\scriptsize ion}}$}}
\newcommand{\Esurf}{\mbox{$E_{\mbox{\scriptsize ss}}$}}
\newcommand{\Eimf}{\mbox{$E_{\mbox{\scriptsize im}}^{(1)}$}}
\newcommand{\Eims}{\mbox{$E_{\mbox{\scriptsize im}}^{(2)}$}}
\newcommand{\vnor}{\mbox{$v_{\mbox{\scriptsize nor}}$}}
\newcommand{\vpar}{\mbox{$v_{\mbox{\scriptsize par}}$}}

\newcommand{\eq}[1]{Eq.~(\ref{#1})}
\begin{document}

\title{Ion survival in grazing collisions of H$^-$ with vicinal nanosurfaces probes subband electronic structures}%

\author{John Shaw}
\email[]{jshaw1969@gmail.com}
\affiliation{%
Department of Natural Sciences, D.L.\ Hubbard Center for Innovation,
Northwest Missouri State University, Maryville, Missouri 64468, USA}

\author{David Monismith} 
\affiliation{%
559th Software Maintenance Squadron, Tinker AFB, Oklahoma, USA}

\author{Yixao Zhang}
\affiliation{%
Department of Natural Sciences, D.L.\ Hubbard Center for Innovation,
Northwest Missouri State University, Maryville, Missouri 64468, USA}
\altaffiliation{Cornell University, New York, USA}

\author{Danielle Doerr} 
\affiliation{%
Department of Natural Sciences, D.L.\ Hubbard Center for Innovation,
Northwest Missouri State University, Maryville, Missouri 64468, USA}
\altaffiliation{}

\author{Himadri S. Chakraborty}
\email[]{himadri@nwmissouri.edu}
\affiliation{%
Department of Natural Sciences, D.L.\ Hubbard Center for Innovation,
Northwest Missouri State University, Maryville, Missouri 64468, USA}

\date{\today}


%%%%%%%%%%%%%%%%% END OF PREAMBLE %%%%%%%%%%%%%%%%

\begin{abstract}
We study the electron dynamics at a monocrystalline Pd(111) surface with stepped vicinal nanostructures modeled in a simple Kronig-Penney scheme. The unoccupied bands of the surface are resonantly excited \textit{via} the resonant charge transfer (RCT) interaction of the surface with a hydrogen anion reflected at grazing angles. The interaction dynamics is simulated numerically in a quantum mechanical wave packet propagation approach. Visualization of the wave packet density shows that, when the electron is transferred to the metal, the surface and image subband states are the most likely locations of the electron as it evolves through the superlattice. The survival probability of the interacting ion exhibits strong modulations as a function of the vicinal-terrace size and shows peaks at those energies that access the image state subband dispersions.  A simple square well model producing standing waves between the steps on the surface suggests the application of such ion-scattering at shallow angles to map electronic substructures in vicinal surfaces. The work also serves as the first proof-of-principle in the utility of our computational method to address, via RCT, surfaces with nanometric patterns. 
\end{abstract}
\maketitle

\section{Introduction}

A vicinal surface consists of a thin-film substrate for which the normal to the surface deviates slightly from a major crystallographic axis. Vicinal surfaces of regular arrays of linear steps, prepared by cutting and polishing a single crystal followed by ultra-high vacuum methods like ion sputtering, are the simplest models of lateral nanostructures on the surface. These high Miller index surfaces are thought to mimic more closely rough regions of industrial surfaces. Such surfaces can be critical for their catalytic properties, without losing a well-defined lattice periodicity \cite{pratt}. The electronic motions in these surfaces can be particularly fascinating because of the possibility of electron scattering at step edges to induce one- or two-dimensional confinement. 

Many metallic vicinal surfaces are of eminent interest due to the presence of a Shockley surface state and image states, respectively, on and above the corresponding flat surface. These states arise from a broken translational symmetry along the surface normal which confines the state in this direction by the crystal band gap. It is therefore expected that vicinal nano-stepping will modify the electronic properties of the surface and image state \textit{via} the superlattice effects. Significant alterations of the electronic properties of surface electrons have been detected for Cu and Au vicinals by angle-resolved photoemissions in momentum space with synchrotron radiation \cite{mugarza, ortega} and scanning tunneling microscopy in the real space \cite{didiot, suzuki, bolz}.  Additionally, ultraviolet photoelectron spectroscopy has indicated unique two-dimensional Shockley surface states on (332) and (221) vicinals of Cu \cite{baumberger, greber}. Furthermore, quantized states are known to form in front of surfaces due to the polarizing image-interaction of an external electron. The investigation of these image states is a powerful tool for probing a variety of physical and chemical phenomena on the nanometer scale. In particular, for metallic vicinals both the confinement and superlattice effects can produce significant image-band splittings and anticrossings from lateral back scatterings at the step edges as predicted theoretically within a static impenetrable surface model \cite{lei}. Therefore, to gain insights in the electronic motions and excitations in such materials, it becomes essential to devise and implement an appropriate theoretical methodology. This is necessary to describe and simulate the details of processes in the band structure altered by modifications on the surface. One theoretically amenable way to do this is to simulate the motion of active electrons in the presence of an impinging negative ion. These results can be tested experimentally and can be utilized to guide future theoretical modeling. 

The charge transfer interaction dynamics of an atomic or molecular ion with a surface is highly sensitive to the surface electronic band structure. From a fundamental science perspective, the understanding of the dynamics is useful to describe phenomena, namely, scattering, sputtering, adsorption, and molecular dissociation~\cite{rabalais03}. Technologically, this process is a crucial intermediate step in (i) analysis, characterization, and manipulation of surfaces~\cite{stout00}, (ii) micro-fabrication based on reactive ion etching and ion lithography~\cite{campbell01}, and (iii) semiconductor miniaturization and the production of self-assembled nanodevices~\cite{korkin07}. In recent past, electron transfer between various atomic species with surface containing nanosystems has become a topic of particular interest. Instances include probing effects of the nanosystem's size and shape to determine their electronic structures~\cite{canario03}. 

The energy conserving transfer of a single electron, the resonant charge transfer (RCT), occurs when the shift of the ion affinity-level enables the transfer to (from) an unoccupied (occupied) resonant state of the substrate. The RCT process in ion-scattering has been the focus of a number of experimental studies on mono- and polycrystalline metal surfaces~\cite{bahrim, yang, hecht, guillemot, sanchez}. A wave packet propagation method was used to access RCT dynamics between excited states of Na nanoislands on the Cu(111) substrate~\cite{hakala07}. Also, recent theoretical research studied the RCT interactions of negative ions with nanoisland films~\cite{gainullin15}. In previous years, we undertook a full quantum mechanical wave packet propagation approach to perform detailed theoretical studies of RCT ion-scattering and associated ion-neutralization processes in the light of altering crystallographic properties of atomically flat (i.e. low Miller index) surfaces \cite{chak70, chak69, chak241, schmitz}, with success in agreeing with measurements~\cite{chak69}. For surfaces with periodic arrays of terraces, the confinement driven reflections of electrons from steps can further structure the free electron dispersion into subband dispersion to enrich the RCT process.  

Therefore, in the present study, the same techniques are applied to vicinally stepped Pd(111) surfaces to calculate the hydrogen negative-ion ($H^-$) survival probability. Even though we use a rather simple model for the vicinal corrugation, the previous utilization of this model in interpreting angle-resolved photoemission measurements on such surfaces to access surface electronic states~\cite{mugarza} provides some confidence in probing, at least qualitatively, the ion-vicinal RCT process in a fully quantum mechanical time propagation treatment. The calculations are done for different $H^-$ collision velocities parallel to the surface at shallow incident angles as well as for different distances between the steps on the surface. The electron wave packet probability density was calculated at all points in space at each time interval. Visualization of this data showed that, when the electron transferred from the ion to the metal, it most likely transferred to both the surface and image superlattice states when the ions approach velocity perpendicular to the surface is slow enough. For a given distance between the steps on the surface, the ion survival probability shows peaks at certain velocities. It is shown that these peaks appear when the kinetic energy of the electron transferred to an image state matches the subband dispersions supported by the periodic vicinal steps. The result suggest a possibility of accessing superlattice band structures \textit{via} anion-scattering in experiments. Atomic units (a.u.) are used in the description of the work, unless mentioned otherwise.

\section{Essentials of the Method}

\subsection{Surface Model}

%%%%%%%%
\begin{figure}[h!]
\includegraphics[width=8.3cm]{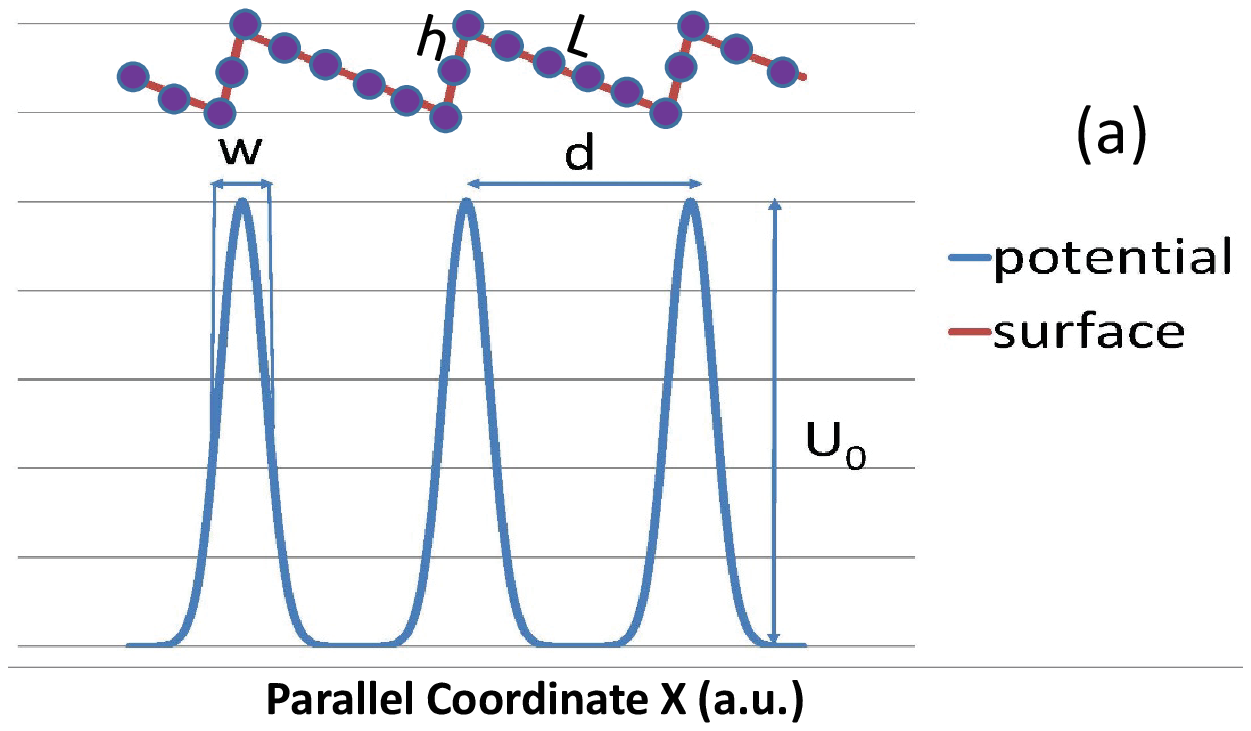}

\centering
\includegraphics[width=8.3cm]{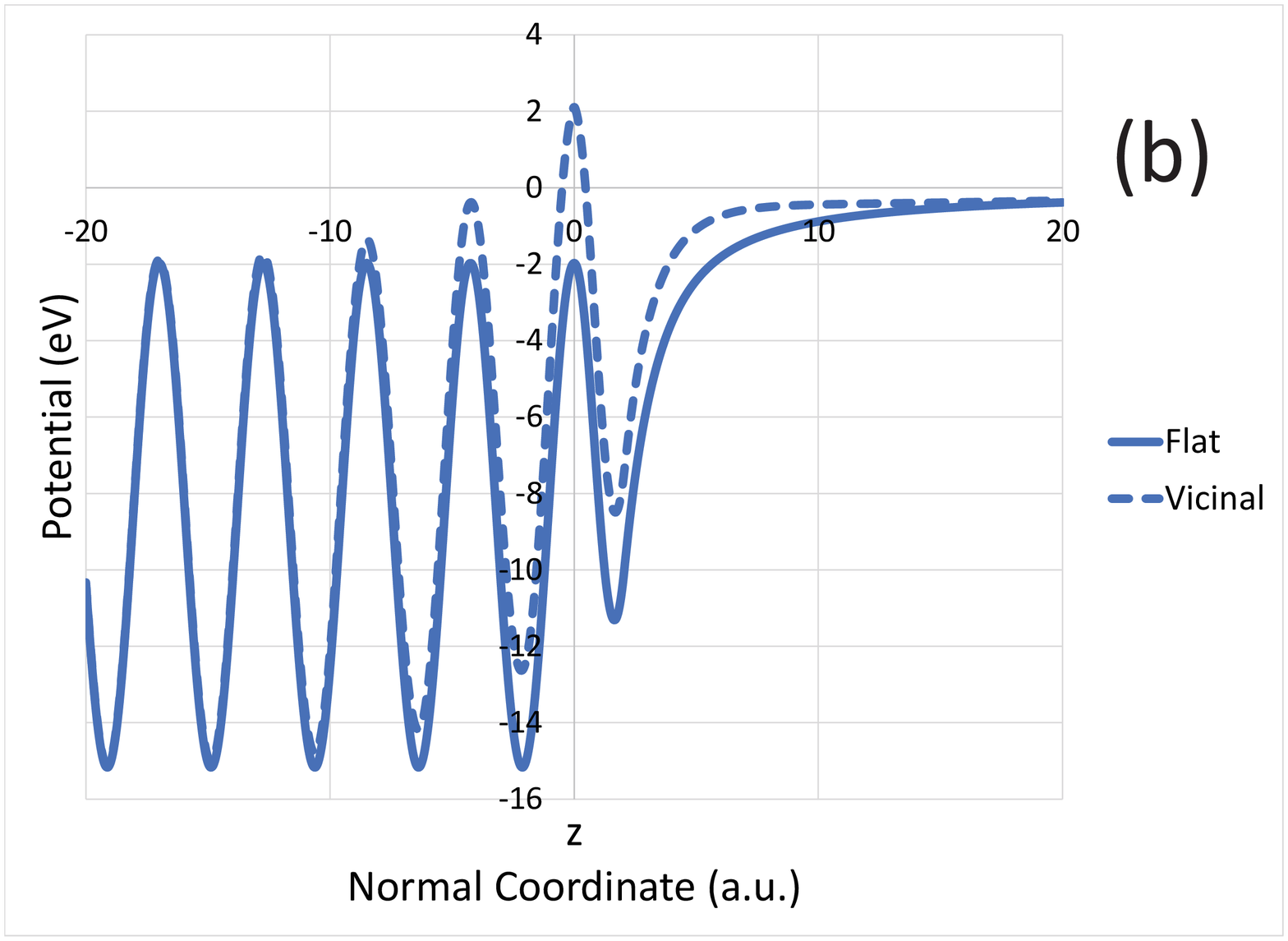}

\caption{(Color online) (a) The one-dimensional Kronig-Penney potential~\cite{mugarza} and the vicinal surface it models with terrace width ($L$) and step height ($h$). (b) The solid line is the one-dimensional pseudopotential of flat Pd(111)~\cite{chulkov}, while the dashed line shows the addition of the potential, as in (a), of the vicinal surface at one of its peaks, scaled by a factor of 2.8 and duly attenuated (see text). This curve is a $z$-section of Fig.\ 2}
\label{fig:fig1}
\end{figure}

\begin{figure*}
\centering
\includegraphics[width=17cm]{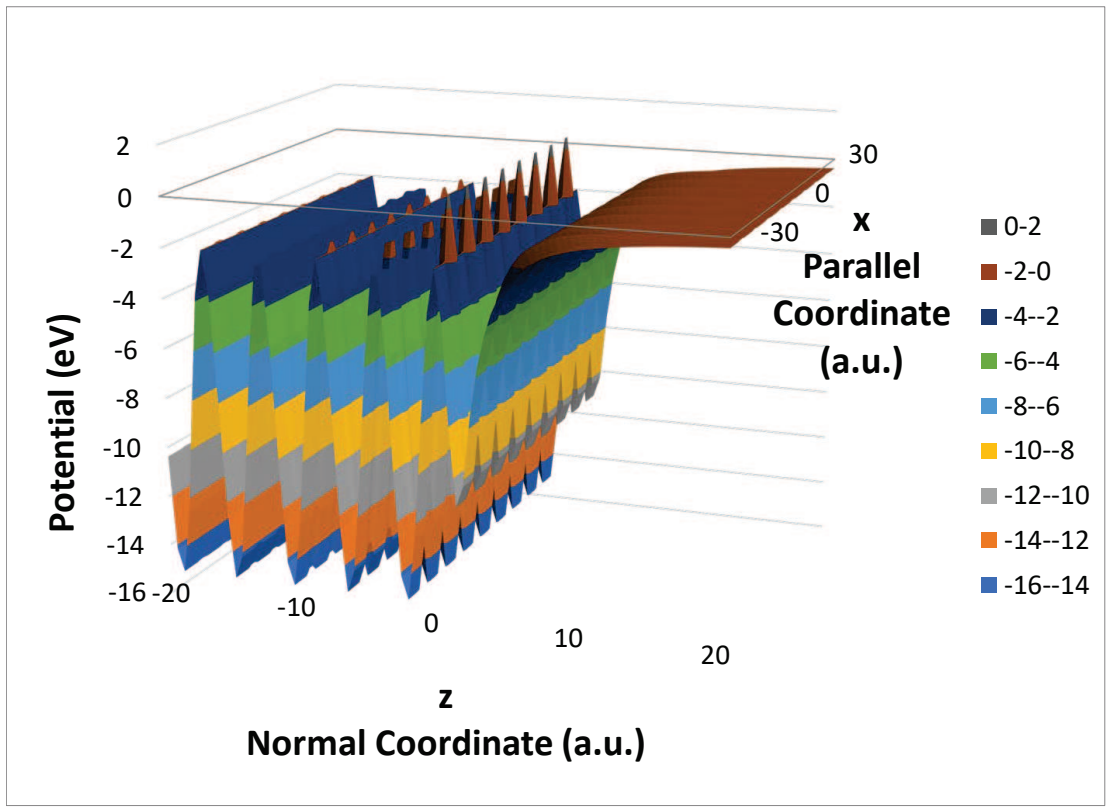}
\caption{(Color online) Schematic of two-dimensional vicinal stepped potential on flat Pd(111) with $d$ being five atomic layers developed using Kroenig-Penny (KP) potential in Fig.\ 1(a) being super-imposed on one-dimensional pseudopotential of Pd(111)~\cite{chulkov} in Fig.\ 1(b). The KP potential is attenuated going far from the precursor flat surface both towards the bulk and the vacuum. To aid visualization, the peaks of the KP potential are scaled by a factor of 2.8 in the figure. Note that the maximum vicinal peaks occur at $z=0$.}
\label{fig:fig2}
\end{figure*}
%%%%%%%%
Due to the repulsive step-step interactions, vicinal corrugations generally appear regularly spaced~\cite{swamy99}. The size of the terrace is determined by the terrace width (vicinal miscut) $L$ and height $h$ of the step. We made use of a two-dimensional model of the metal surface which includes the ion approach direction ($z$) normal to the primordial flat surface Pd(111) and only one direction ($x$) on the flat surface along the vicinal steps. We used the Kronig-Penney (KP) potential to mimic the periodic potential array formed by the step superlattice in the $x$-coordinate shown in Fig.\ 1(a). The steps in the metal surface are mimicked by the peaks in the KP potential with height $U_0$ and width $w$, and the step separation $d$ (distance between adjacent steps). This model is used as it was successful at describing the experimental photoemission spectra for stepped vicinal metal surfaces \cite{mugarza}. Experiments used vicinal surfaces with the step height of a single atomic layer cut. When fitting the data to the KP model, the largest value of the product $U_0w$ was found to be 0.054 a.u.\ in Ref.\ [\onlinecite{mugarza}]. Therefore, we chose $U_0$ = 0.054 a.u.\ with $w$ = 1 a.u.\ for the results presented in this article. We note that there is no exact correspondence between $U_0$ and $h$. This is because our model does not represent vicinal steps at the atomistic level which needs a full 3D structure calculation. Rather, in the spirit of Ref.\ [\onlinecite{mugarza}], the model mimics the effect of vicinal steps using a flat surface with a potential array for which the strength $U_0w$ correlates the electrostatic strength of a step. We should also note, as discussed in Ref.\ [\onlinecite{mugarza}], that approximating the periodic potential as a Dirac $\delta$-array $\sum _n U_0w \delta(x-nd)$, or other combinations of $U_0$ and $w$ producing the same barrier strength $U_0w$ will not change the model. Indeed, for all our calculations a different combination, namely $U_0$ = 0.027 a.u.\ and $w$ = 2 a.u., fully reproduced the results. We emphasize that in the absence of any parametric form of the vicinal surface potential derived from ab\,initio calculations, we use a scheme of combining this one-dimensional array potential with a known parametric representation based on an ab\,initio flat surface potential as described below.

\subsection{Wave packet propagation}

The details of the propagation methodology are given in Ref. \cite{chak70}. The time-dependent electronic wave packet $\Phi\left(\vec{r},t;D\right)$ for the ion-surface combined system is a solution of the time-dependent Schr\"{o}dinger equation
\begin{equation}\label{tdse}
i{\hbar}\frac{\partial}{\partial t}\Phi\left(\vec{r},t;D\right) = H\Phi\left(\vec{r},t;D\right),
\end{equation}
with the general form of the Hamiltonian as
\begin{equation}\label{hamil}
H = -\frac{1}{2}\frac{d^{2}}{dz^{2}}-\frac{1}{2}\frac{d^{2}}{dx^{2}}+V_{\mbox{\scriptsize vi-surf}} (x,z) + V_{\mbox{\scriptsize ion}} (x,z)
\end{equation}
where $D(t)$ is the dynamically changing perpendicular distance between the $z=0$ line (see below) of fixed-in-space metal surface and the ion moving along a trajectory.  The potential, $V_{\mbox{\scriptsize vi-surf}}$, of the vicinal surface will have two components: (i) A one-dimensional potential model in $z$ obtained from pseudopotential local-density-approximation calculations for simple and noble metal surfaces~\cite{chulkov} that represents the flat precursor substrate. This is the model, with free electron motion in the $x$ direction, that was employed in our previous work with flat surfaces \cite{chak70, chak69, chak241, schmitz}. A graph of this potential appears in Fig.\ 1(b). The lattice points of the topmost layer of the substrate is taken at $z=0$ so the peaks in the potential for $z<0$ are the centers of the atomic layers going into the bulk and separated by the interatomic lattice spacing. (ii) Superimposed on (i) is the KP potential model described above that mimics the regular vicinal array of terraces $L$ on the surface along $x$. To limit the vicinal effect far from the surface, the KP potential is attenuated exponentially in both the positive (vacuum side) and negative (bulk side) $z$-direction. A curve representing a section of $V_{\mbox{\scriptsize vi-surf}}$ along $z$-direction through a vicinal peak is also included in Fig.\ 1(b). Fig.\ 2, on the other hand, provides an illustration of the full 2D $V_{\mbox{\scriptsize vi-surf}}$ used in our simulation. The peaks of the KP potential, as seen in both Fig.\ 1(b) and Fig.\ 2, are set at their maximum (unattenuated) value at $z=0$. This must be the case if the atomic layer positions of the precursor flat surface are not to be altered by the vicinal potential. Therefore, the top atomic layer used in the flat Pd(111) potential in Fig.\ 1(b) defines $z=0$. Smaller peaks are barely visible in Fig.\ 2 on the top of the potential for the next atomic layer inside the surface which is due to the effect of the exponential attenuation. Small peaks can also be seen at the bottom of the dips. The same attenuation is present for $z>0$ but is not as clearly visible in Fig.\ 2.

The $H^-$ ion will be described by a single-electron model potential, $V_{\mbox{\scriptsize ion}}$, which includes the interaction of a polarizable hydrogen core with the active electron \cite{ermoshin}. However, we employ an appropriately re-parametrized version of this potential~\cite{chak70}, as was also applied in our other publications~\cite{chak69, chak241, schmitz}. This form is commensurate with our two-dimensional propagation scheme and produces the correct ion affinity level energy \Eion = 0.0275 a.u. (0.75 eV).

The propagation by one time step $\Delta t$ will yield
\begin{equation}\label{propa}
\Phi(\vec{r},t+\Delta t;D) = exp[iH(D)\Delta t]\Phi(\vec{r},t;D)
\end{equation}
where the asymptotic initial packet $\Phi_{\mbox{\scriptsize ion}}(\vec{r},t=0,D=\infty)$ is the unperturbed $H^-$ wave function $\Phi_{\mbox{\scriptsize ion}}(\vec{r},D)$. The ion-survival amplitude, or autocorrelation, is then calculated by the overlap
\begin{equation}\label{auto}
A(t) = \left\langle \Phi(\vec{r},t)|\Phi_{\mbox{\scriptsize ion}}(\vec{r})\right\rangle.
\end{equation}
We employ the split-operator Crank-Nicholson propagation method in conjunction with the unitary and unconditionally stable Cayley scheme to evaluate $\Phi(\vec{r},t;D)$ in successive time steps \cite{chak70,press}. Obviously, the propagation limits the motion of the active electron to the scattering plane of the ion.

We will assume that when the ion reflects from the surface, the angle of reflection is the same as the angle of incidence measured from the flat substrate plane. The Hydrogen ion impinges on the surface at shallow angles with respect to $x$. Inputs to the calculation are the component of ion velocity normal ($z$) to the surface, $\vnor$, and the component of velocity parallel ($x$) to the surface, $\vpar$. The computer program aims the ion at a point halfway between two steps as well as directly on a step. The incident ion will decelerate along the $z$ direction, close to the surface, due to the net repulsive interaction between the ion core and the surface atoms while it will stay constant in $\vpar$. For a given initial velocity, we will simulate a classical trajectory based on Biersack-Ziegler (BZ) interatomic potentials to model the repulsion \cite{chak70, biersack}. The slowdown of $\vnor$ to zero at the point of closest approach (the turning point) and its subsequent gradual regaining of the initial speed at the initial distance while reflecting symmetrically at constant $\vpar$ makes the resulting trajectory somewhat parabolic. This ionic motion is then incorporated in the propagation by adding the translational phase $(\vnor z + \vpar x + v^2t/2)$~\cite{chak70} as well as by shifting the center of the ion potential in \eq{hamil} to follow the BZ trajectory that corresponds to evolving $D(t)$. We note that to formulate the trajectory we calculated the BZ potential as if the surface was flat at $z =0$. But this limitation of our BZ trajectory, blind to vicinal shapes, is not expected to qualitatively affect the main results which should predominantly depend on the RCT process.

We remark that in the grazing scattering, the H$^-$ neutralization on a palladium surface (which is different from cation neutralization that has a strong capture rate from metals Fermi sea) can avoid direct consequence of the so called parallel velocity effect (which causes a shift of the Fermi sphere~\cite{borisov96}) over a good range of $\vpar$ due to the following reason. The Pd Fermi energy from the bottom of the valence band is about 0.26 a.u.~\cite{mueller70} which corresponds to the magnitude of Fermi wave vector of about $k_f=$ 0.85 a.u. after accounting in a 41\% raise in the electron effective mass~\cite{mueller70}. This implies that the observed Fermi energy $(k_f-\vpar)^2/2$ from the ion's moving frame~\cite{winter91} may not influence the energy range of the current RCT process [which is very close to the Pd(111) image state energies as discussed in the following section] at least, if not farther, in the range up to about $\vpar = 0.5$ a.u. within which strong structures in the ion survival occur (see Fig.\,3).  

In our simulations, the distance of closest approach to the substrate will be kept fixed. This is reasonable as the initial value of $\vnor$ at $z=20$ a.u.\ will be constant at a small value 0.03 a.u.\ and thereby largely omits the variation of the dynamics as a function of $\vnor$. We are interested in changes of survival probability due to the surface vicinal structure by varying $\vpar$. Long after the ion's reflection, the final ion-survival probability will be obtained by
\begin{equation}\label{surv}
P=\lim_{t\rightarrow\infty}\left|A\left(t\right)\right|^{2},
\end{equation}
which corresponds to fractions of the survived incoming ions that an experiment can measure~\cite{guillemot}.

To calculate the final ion-survival probability, the computer program calculates the electron wave packet density at all points in space at each time interval. This data was used to produce detailed animations of the changing electron wave packet density with time. Though the initial and final value of $z$ was fixed at 20 a.u., the initial and final values of $x$ varied with $\vpar$. Due to a small $\vnor$ value and many values of relatively fast $\vpar$, the ion's flyby was nearly grazing the surface. Consequently, the size of the surface, $|x_{\mbox{\scriptsize final}} - x_{\mbox{\scriptsize initial}}|$, entered in to the propagation was large, resulting in the execution time of the computer program to be very long.  As a result, thread-based parallel computing using OpenMP was employed. To accumulate the amount of data we used, parameter sweeps were done with the program on the Stampede supercomputer at the University of Texas at Austin. More than 500 survival probabilities were calculated in this way for stepped Pd(111) surfaces considered in this paper. It should be noted that calculations were done in the 2D model described above and therefore all figures present 2D model results.

\section{Results and Discussions}

\subsection{Ion survival}

Hydrogen ion survival probabilities on vicinally stepped Pd(111), after the ion returns to the initial $z$ = 20 a.u.\ $(t\rightarrow\infty)$, were calculated for parallel velocities, $\vpar$, ranging from 0.2 to 1.0 a.u.\ in steps of 0.05 a.u. This is equivalent to a range of ion scattering angles from 8.53 to 1.72 degrees with respect to the (111) direction. The survival probabilities were calculated for the distance between steps ($d$) of 5, 7, 9, 11 and 13 lattice spacings ($a_s$), where $a_s$ = 4.25 a.u.\ for Pd(111). The results of the hydrogen ion survival probability are shown in Figs.\ 3 both for the ion heading toward the center of a terrace (solid line) and directly toward a step (dashed line). The graphs were fit to the calculated data points using a cubic spline. The ion survival result, labeled ``F'' in the legend, in front of the flat Pd(111) under the same kinematic conditions is also shown for comparison. As seen, the vicinal steps cause significant modulations in the ion survival probability, whereas the flat Pd(111) result is smooth. 

Note that for $d= 5a_s$ in Fig.\ 3(a) the step-induced modulations are the same whether the ion approaches the center of a terrace or a step, except for extremely low values of $\vpar$. As the distance between steps increases, the lowest value of $\vpar$ for which the center-approaching and the step-approaching results merge becomes gradually larger. For instance, as seen in Fig.\ 3(b), for $d = 13a_s$, the widest terrace considered, the curves merge at $\vpar$ = 0.63 a.u.  We also note that for each choice of $d$ both the curves still produce qualitatively similar structures even over the lower $\vpar$ values where their magnitudes do not agree. This trend in the results has an important implication. Since the scattering angles are small, the ion in general interacts with a large number of steps during its grazing flyby populating the vicinal surface superlattice levels. For the shortest $d$ considered, the number of steps interacting with the ion becomes so large that the level-population and subsequent electron-recapture by the ion becomes practically insensitive to the ion's approach site, center or step. As a result, the corresponding ion-survival results are identical. For gradually larger vicinal steps, however, this condition of approach-site independence is barely met at progressively higher $\vpar$ corresponding to lower (more grazing) scattering angles that results in just enough number of steps to be included in the interaction. Thus, as $d$ increases, a corresponding decrease in the grazing angle is needed for a ``perfect'' excitation of vicinal subbands when the site on the surface the ion is heading to does not matter anymore. Consequently, meeting this condition may bring benefit in conducting experiments where one may not need to worry about the vicinal surface location the ion beam aims at.

In principle, the ion survival on a flat surface should be smooth as a function of $\vpar$ for a fixed $\vnor$ owing to the parabolic free electron dispersion of a flat surface. Moreover, for a flat surface with a projected band gap in the direction normal to the surface, like in Pd(111)~\cite{chulkov}, that resists decay in $z$-direction, the survival should be almost steady as a function of $\vpar$. However, our flat Pd(111) surface result in Figs.\ 3 is showing a slow monotonic decrease which is likely because of a numerical artifact due to slightly imperfect boundary absorbers used in the simulation to mimic an infinite surface. It is then expected that this overall background artifact also exists for various vicinal surface results, but that does not alter the main results of modulations in the ion survival arising from the RCT electrons' access to the subband structure. The reason for this is that the error introduced by the boundary absorbers is a systematic error. It will effect all results by the same degree. This will not change the position of the peaks in the survival probability, only their relative amplitudes. 

We also found (not shown here) that if the product $U_0w$ was half as great, that is 0.027 a.u., the survival peaks were in the same location but half as high. This is due to the fact that the electron transmits more easily to the next well for smaller $U_0w$. Specifically, with $U_0w$ being halved, the transmission probability doubles, lowering the survival peaks as it did. However, the fact that the peaks were in the same location indicates that the distance between steps is the only surface structure property that effects the position of the peaks.

It is true that the results in Fig.\ 3 show variations of about $\pm$0.2\% around a small average H$^-$ survival probability (ion fraction) of roughly 2\%. However, experimentally an ion fraction as little as 0.1\% has been measured for flat surfaces within an error range of about $\pm$0.1\%~\cite{guillemot} over similar ion-velocities used in this work. Therefore, we believe that structures in our current predictions should be accessible in the experiment.

A final note concerning the use of a 2D model. It is our purpose to qualitatively understand the main physics happening during the ion-vicinal surface interaction. A 3D model would change the magnitudes of the survival probabilities but would not change the physics we are proposing to account for the structures observed in these survival probabilities.

%%%%%%%%
\begin{figure}[h!]
\includegraphics[width=8.3cm]{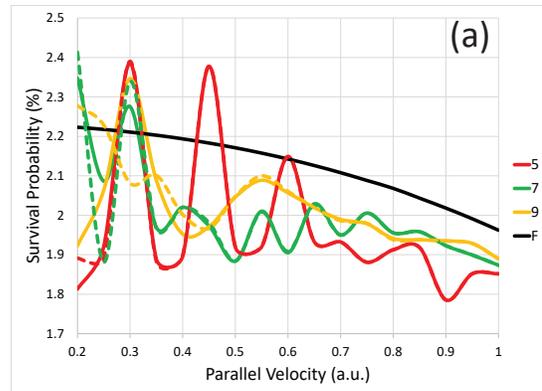}

\centering
\includegraphics[width=8.3cm]{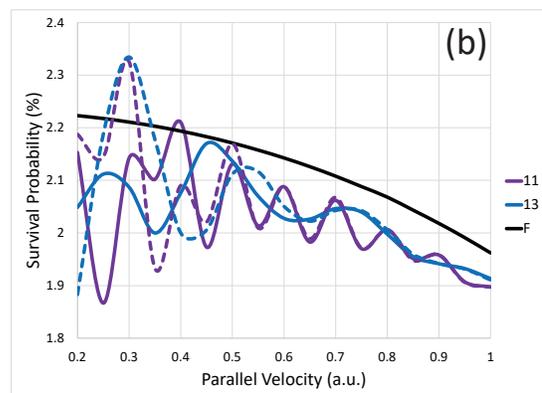}
\caption{(Color online) Hydrogen ion survival probability for one atomic-layer high vicinally stepped Pd(111) as a function of ion parallel velocity ($\vpar$) as $H^-$ approaching the center of a terrace (solid line) and a step (dashed line) for terrace width ($L$) of 5, 7, and 9 lattice spacings (a) and for terrace widths of 11 and 13 lattice spacings (b).}
\label{fig:fig3}
\end{figure}
%%%%%%%%
%%%%%%%%
%\begin{figure*}
%\includegraphics[width=13cm]{Fig3.eps}
%\caption{(Color online) Time-snapshots of the electron wave packet density: (a) for the flat Pd(111) at the time $t$; (b) for the vicinal Pd(111) of $L$ =5$a_s$ at the same instant of (a); (c) the same as (a) but for a 90$^o$-rotated view and at a slightly different time $t'$; (d) same as (c) but at a later time $t'+\Delta t$.} 
%\label{fig:fig3}
%\end{figure*}
%%%%%%%%
%%%%%%%%
\begin{figure*}
\centering
	\begin{minipage}[b]{8.3cm}
	\includegraphics[width=8cm]{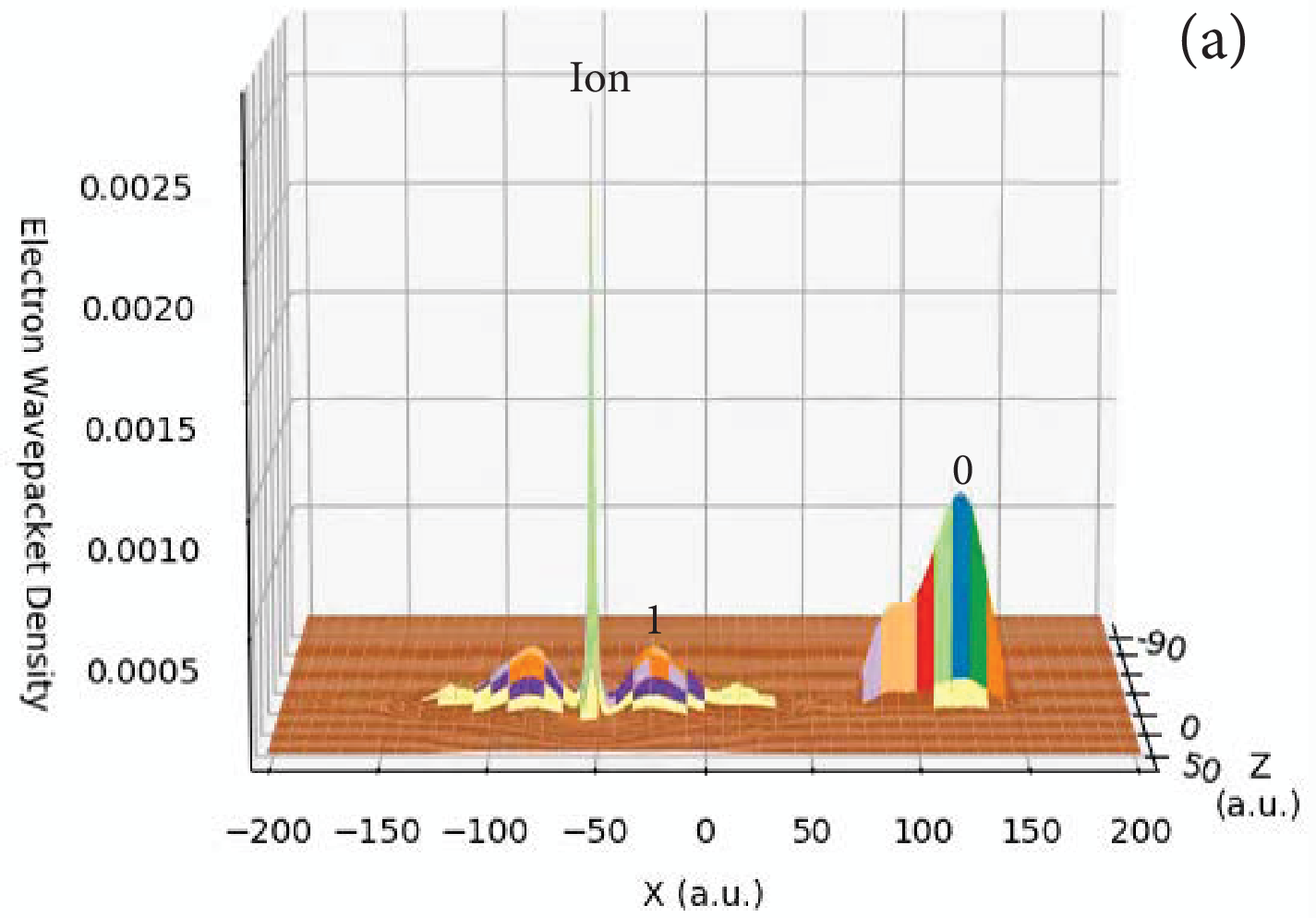}
	\end{minipage}\quad
	\begin{minipage}[b]{8.3cm}
	\includegraphics[width=8cm]{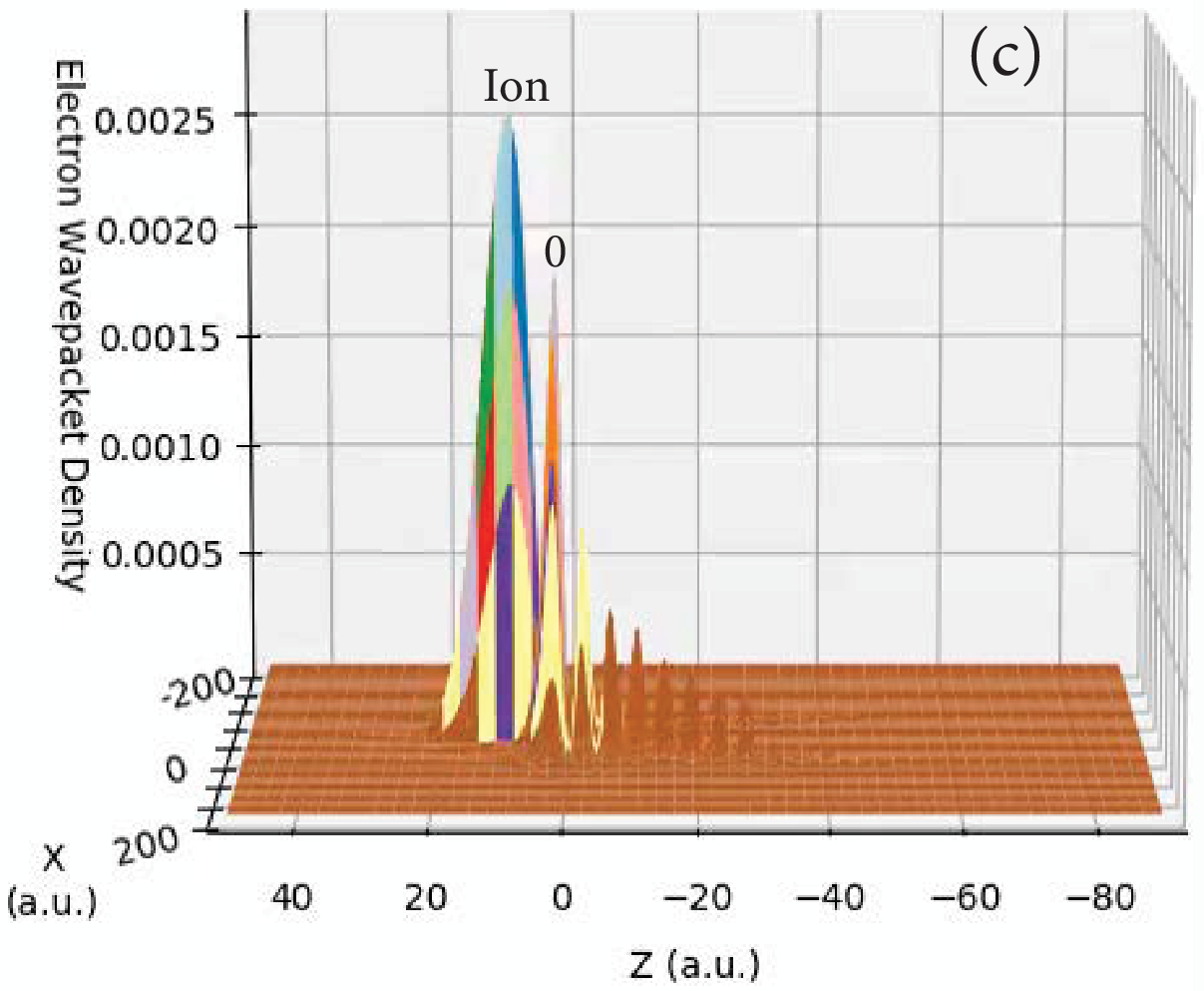}
	\end{minipage}\quad
	\begin{minipage}[b]{8.3cm}
	\includegraphics[width=8cm]{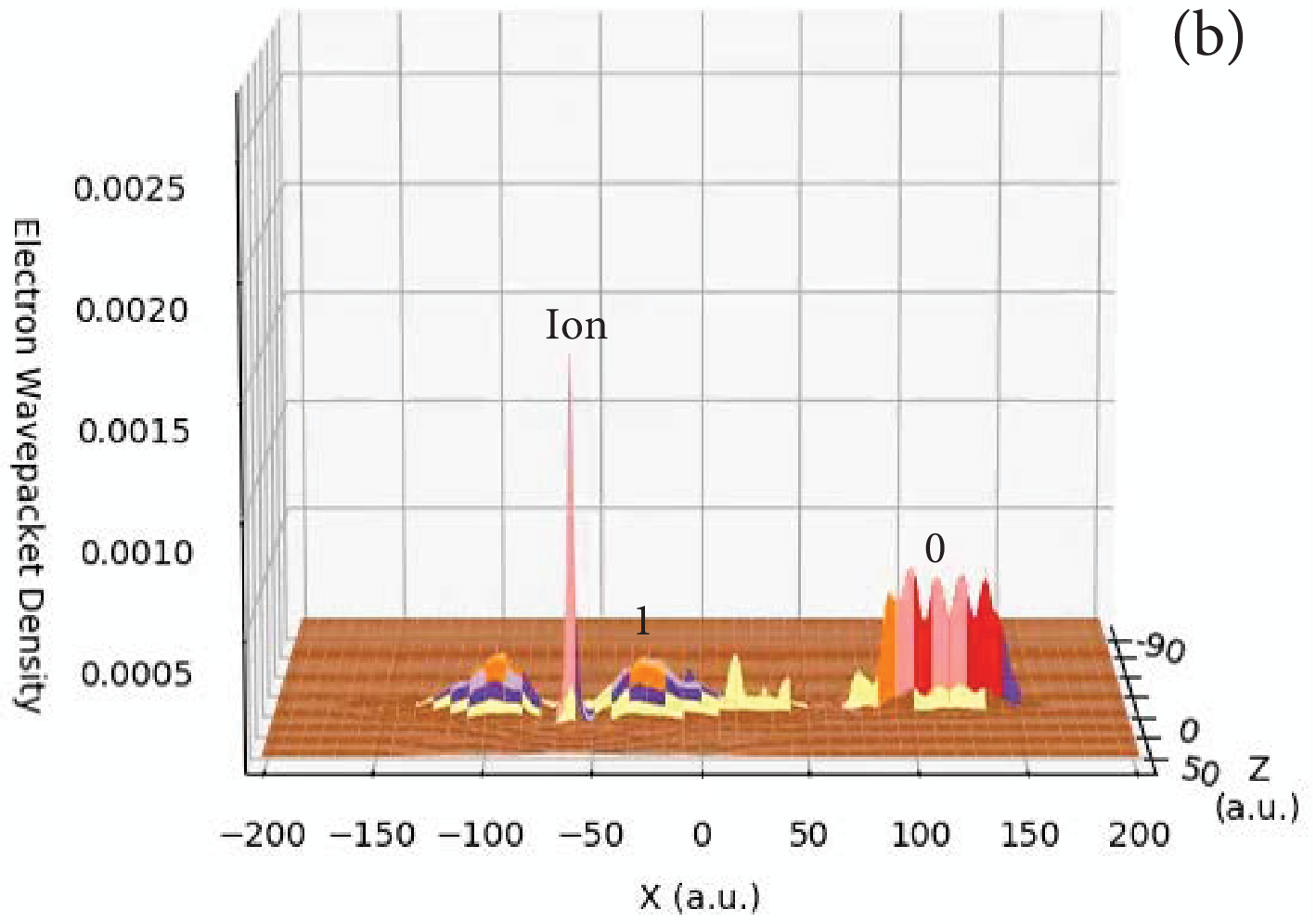}
	\end{minipage}\quad
	\begin{minipage}[b]{8.3cm}
	\includegraphics[width=8cm]{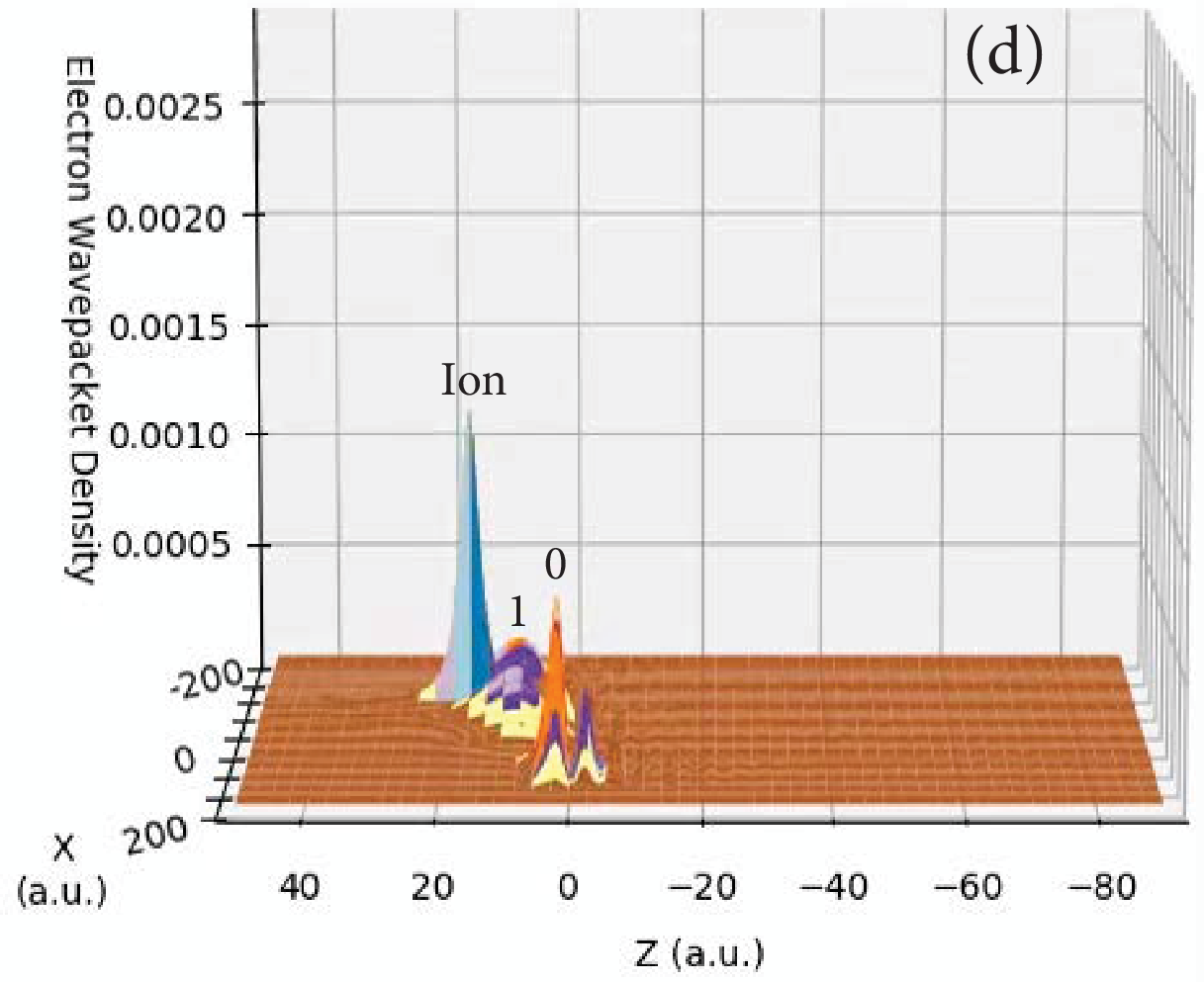}
	\end{minipage}
\caption{(Color online) Time-snapshots of the electron wave packet density: (a) for the flat Pd(111) at the time $t$; (b) for the vicinal Pd(111) of $d$ =5$a_s$ at the same instant of (a); (c) for the vicinal Pd(111) of $d$ =5$a_s$ for a 90$^o$-rotated view at a different time $t'$ than in (b); (d) same as (c) but at a later time $t'+\Delta t$. The electron wave packet density is a dimensionless fraction which is normalized to unity over all space.}
\label{fig:fig4}
\end{figure*}
%%%%%%%%

\subsection{Wave packet density dynamics}

The projected band gap of Pd(111) in the $z$ direction exists across the vacuum level energy $E_v$ = 0 and extends from its lower edge -0.163 a.u.\ to the upper edge 0.0794 a.u.~\cite{schmitz,chulkov}. There also exists a Shockley surface state localized inside the gap at \Esurf = -0.152 a.u.\ and Rydberg series of image states converging to the vacuum level with the energies of the first two image states being \Eimf = -0.0202 a.u.\ and \Eims = -0.00625 a.u. It is therefore expected that the incoming ion at its closest approach to the surface will resonantly populate both the surface and image state bands~\cite{schmitz}. The ion survival amplitude, \eq{auto}, dynamically evolves as the ion approaches the surface. When the ion is close to the surface, there are one or more positions for which the amplitude is found zero. This means that at close distances the electron transfers back and forth between the ion and the metal. Therefore, it is important to probe the RCT dynamics involving the metal states that the ion populates. The vicinal texture is expected to induce superimposed subband modulations to produce survival peaks in Fig.\ 3. Therefore, to understand the surface state and image state RCT dynamics in an uncluttered manner, the wave packet densities on flat Pd(111) are analyzed in Fig.\ 4(a) and the density for vicinal Pd(111) is addressed in Fig.\ 4(b), 4(c) and 4(d).

Electrons transferring from the ion to a much deeper surface state of Pd(111) will acquire large kinetic energy resulting in a maximum speed $\sqrt{2(\Eion-\Esurf)} \approx 0.5$ a.u.\ along $x$-axis on the surface to either direction from the closest approach; we neglect the energy from very slow $\vnor$ = 0.03 a.u.\ throughout this article and assume strong resistance to decay into the bulk due to the band gap. Indeed, animations produced by the wave packet probability density as a function of time confirm this later assumption. We present some time snapshots from our animation for $\vpar$ = 0.2 a.u.\ and $d = 5a_s$ in Figs.\ 4 which show the electron wave packet density as a function of $x$ and $z$. The $x$-axis is across the page and the $z$-axis is into the page for Fig.\ 4(a) for flat Pd(111). For this figure, the ion starts at the right side of the graph and heads into the paper toward a terrace center and to the left, approaching closest to the surface at $x$ = 0. Thus, since $\vpar$ is toward left, the resonantly transferred electrons to the surface state will move in a speed of $0.5+\vpar$ along the negative $x$-axis and in a slower speed of $0.5-\vpar$ along the positive $x$-axis. In Fig.\ 4(a), the narrow peak in the middle is the location of the ion. But the faster moving surface state density to the left is mostly absorbed by the numerical absorber at the grid-edge built into the code to simulate infinite surface, while its slower moving counterpart to the right labeled 0 is still visible. Consequently, the fast moving surface state density, rapidly separating from the ion as seen in Fig.\ 4(a), makes it impossible for the depleted ion to recapture electrons. Therefore, while the propagation of the full wave packet uncovers the evolution of the populated surface band, this band, forbidding recapture, plays virtually no role in determining the structures in the final ion survival, except that it drains the ion significantly. On the other hand, image state energies being very close to \Eion will have small enough excess energy to enable the populated image densities to stay closer to the moving ion. As seen in Fig.\ 4(a), of the two density components of image electrons closely flanking the ion the feature on the right is labeled as 1. In fact, it is known that a largely slowed down ion near the surface will mimic adiabatic type interactions causing the affinity level and the surface-state level to repel each other~\cite{chak69,chak70}. This will further increase and decrease, respectively, the ion-surface state and ion-image states transition energies to more favor the dynamics described above.  For a surface with vicinal steps, the RCT electrons from ion to surface will therefore evolve through the subband structures of these states. Fig.\ 4(b), for vicinally textured Pd(111) but at the same instant as of Fig.\ 4(a), confirms such structured density distributions. These structured distributions possess peaks and valleys that one would expect from an interference pattern produced by probability waves reflecting back and forth between the vicinal steps of the surface.

Figures 4(c) and 4(d) show the view along the $z$-axis for vicinal Pd(111). The ion and peaks are labeled in these figures as 0 and 1 as well. In Fig.\ 4(c), the tall peak in front of the surface is the location of the ion. The peak, labeled 0, also in front of the surface, occurs due to the surface state. The probability of finding the electron in this state is the greatest of all other states. Note the decay of the surface state density amplitudes into the bulk, which is a direct consequence of the projected band gap dispersion along $z$ direction, ensuring its localized nature. In Fig.\ 4(d), which is at a later time than Fig.\ 4(c), the surface state feature is seen to have moved out along the $x$-axis whereas a smaller peak from image state densities, labeled 1, remains close to the ion. These are the same peaks seen in Figs.\ 4(a) and 4(b) from a different perspective. We further note that Figs.\ 4(c) and 4(d) appear relatively smoother, since the density structures, as in Fig.\ 4(b),  due to vicinal patterns are only visible along the $x$ direction. In any case, these visualizations of evolving wave packet density thus suggest that while the RCT transferred electrons to the surface of vicinal steps will excite both surface and image state subbands, the image densities, by following the ion from the closest proximity and thereby having a large wave function overlap, will mainly influence the recapture by the outgoing ion and hence its final survival probability~\cite{schmitz,chak241}. Consequently, even though bulk of the electrons are decaying \textit{via} the surface state subband, the modulation in ion's survival as a function of its speed, as seen in Figs.\ 3, will result from the ion's interaction with the image subband as we discuss below. 

%%%%%%%%
\begin{figure}
\centering
\includegraphics[width=8.3cm]{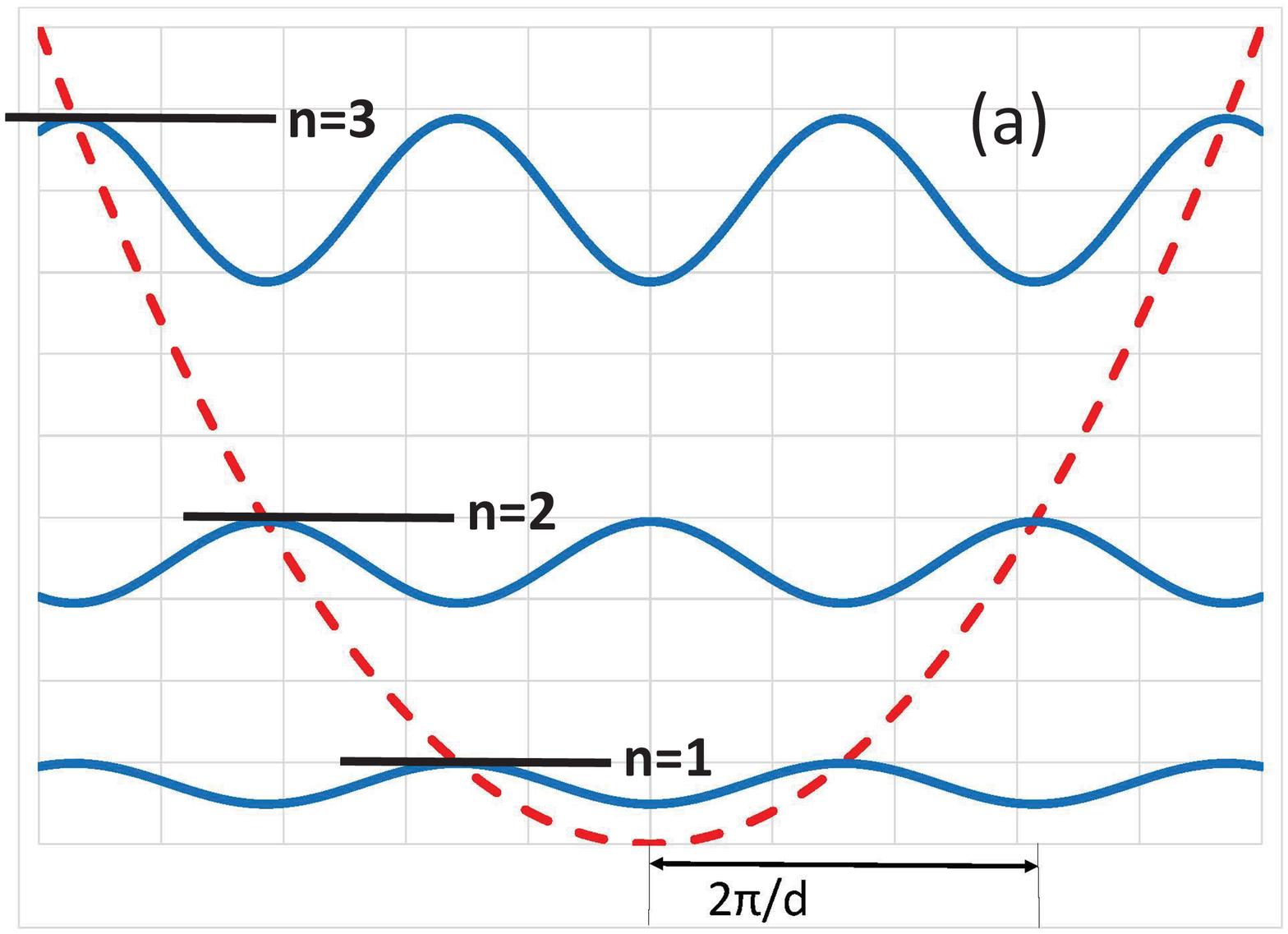}

\centering
\includegraphics[width=8.3cm]{Fig5b.eps}
\caption{(Color online) (a) A schematic of Kronig-Penney superlattice subband dispersions with finite potential barrier representing the vicinal steps. The reciprocal superlattice vector $2\pi/d$ of the subbands and flat energy levels that subbands become for the infinite square well are shown. The corresponding flat surface dispersion is also sketched. (b) Comparison of the ion survival peak positions in the parallel velocity scale for various values of step separation $d$ with an analytic square well potential model.}
\label{fig:fig5}
\end{figure}
%%%%%%%%

\subsection{Superlattice states from lateral confinement}

The electrons in flat Pd(111) exhibit free electron dispersion. The parallel velocity of the moving ion along grazing trajectories predominantly follows the dispersion energy of the RCT electrons into the image states in a parabolic form $(\vpar)^2/2$, since the transition energy from the affinity to, for instance, the first image level is a miniscule $\Eion-\Eimf$ = 0.0073 a.u., and $\vnor$ is chosen to be too small in this work. This would ensure a monotonic recapture rate by the ion and therefore a rather structureless ion survival probability appears as a function of $\vpar$ for flat Pd(111), as is seen in Figs.\ 3. However, the periodic array of potential barriers with finite $U_ow$ and period step separation $d$, representing the vicinal terraces on Pd(111), will induce both reflections and transmissions, respectively from and through the steps, of electrons propagating along the surface. This will give rise to subbands and even forbidden gaps in $x$-direction with the reciprocal vector $2\pi/d$, which zone-folds the subband~\cite{mugarza}; a schematic of the subband dispersions is also shown in Fig.\ 5(a). It is therefore expected that, roughly, whenever the ion's parallel energy, represented by the broad parabolic dispersion in Fig.\ 5(a) will intersect a subband dispersion curve, a resonance-type condition is reached, and the recapture rate by the ion will increase. This will produce peaks in the survival probability, as our main results show in Fig.\ 3. 

One simple way to verify that these subband states from surface-parallel confinements induce survival peaks is to compare our numerical results with the analytic infinite-barrier square-well model. This is because ultimately in the infinite barrier limit, these subband dispersions will turn into flat quantum levels in Fig.\ 5(a), and therefore can still qualitatively guide in capturing the positions of the survival peaks as we shall show below. For the infinite square well potential, the allowed electron de Broglie wavelengths are given by $\lambda_{n}=2d/n$, where $n$ is a positive non-zero integer we shall call the quantum number and $d$ is the width of the well. From this, it is straightforward to show that the quantized kinetic energy of an electron in the well, in atomic units, is $\left(n \pi/d \sqrt{2}\right)^{2}$. By RCT energy conservation, setting this kinetic energy to be equal to the ion's parallel kinetic energy $(\vpar)^2/2$, {\em plus} the transition energy from the ion level to an image level, one solves for $\vpar$ as 
\begin{equation}\label{sq-well}
\vpar=\sqrt{2\left((n \pi/d \sqrt{2})^{2}+\Eion-E\right)}
\end{equation}
that will give standing electron probability density, where $E$ is an image state energy for the flat surface. All quantities are in atomic units.

\eq{sq-well} is plotted for the first (solid lines) and second (dashed lines) image states in Fig.\ 5(b) for five values of $d$ considered in this work. The values of $\vpar$ for which peaks occur in Fig.\ 3 are also plotted as symbols on this same graph for simulations when the ion aims both at the center of a terrace (solid symbols) and a step (hollow symbols). The vast majority of the symbols fall reasonably close to corresponding lines obtained from the model, guiding the broad underlying physics. There are a few symbols which are less close to a line than others. The level of agreement is, however, reasonably good, given the realities that (i) the ion interacts with a subband dispersion process but not with an ideal well of infinite height and (ii) the image population density spreads over the entire image state Rydberg series inducing some uncertainty in the transition energy. Furthermore, the pattern of the quantum numbers of the peaks is what we expect. When $d$ increases, we expect the quantum number of standing probability wave that will fit this distance to increase. We also expect that for a given velocity, the wavelength does not change. Similarly, as the velocity, $\vpar$, increases, the wavelength will decrease and so we expect the quantum number of standing probability wave that matches this velocity to increase as well. Those are precisely the patterns observed in Fig.\ 5(b). One discrepancy between the model and the numerical results can be noted: For $d$ = $9a_s$ and $13a_s$ some bands are intermittently missing. This is likely due to forbidden gaps in the subband structure that exist for these vicinal widths within the $\vpar$ range considered which, however, is the part of finer details of the band structure. 

To this end, our results indicate that one interesting spectroscopic pathway to probe subband dispersions in vicinal superlattice is RCT studies of negative ions in grazing scattering off stepped metal surfaces. As we show, the simpler approach to that goal is to choose the impact energies and scattering angles in such a way that the ion velocity perpendicular to the flat surface direction stays constant while varying in the parallel direction. This will keep the interaction largely free of alterations from the band structure perpendicular to the precursor flat surface.    

\section{Conclusion}

With the example of Pd(111) vicinally miscut in a selection of terrace sizes within a few nanometers we simulate the dynamics of the active-electron in a hydrogen anion projectile impinging at shallow angles to the surface.  To do so, we used a fully quantum mechanical wave packet propagation methodology. The electron dynamics is visualized in details by animating the wave packet probability density in real time. The results, for the first time,  show structures in the ion survival probability due to the image state subband dispersion introduced by the vicinal superlattice. In the absence of a fully ab\,initio potential, we use a simple but successful model of vicinal stepping. This produces structures in the ion survival as a function of terrace sizes, proving the ability of our wavepacket propagation methodology to study RCT tunneling between anions and nanostructured surfaces for the first time. Even though the calculation is based on one-active-electron model, effects of electron correlations due to occupied valence band can only enhance the recapture rate by impeding decay into the bulk from the Pauli blockade. The effects uncovered can and should be observed in grazing ion spectroscopy within the current laboratory technology, although accessing effects of the azimuthal orientation of the scattering plane will require full 3D simulation. The invariance of the results to the strike location on the surface over the higher parallel velocity range suggests that for a sufficiently grazing flyby aiming the ion beam is likely irrelevant, providing some experimental freedom. Comparisons of results among other metal surfaces with vicinal textures, a part of the outlook of our research, will be published elsewhere.

\begin{acknowledgments}
The authors would like to acknowledge the support received by an Extreme Science and Engineering Discovery Environment (XSEDE) allocation Grant for high performance computation, which is supported by National Science Foundation (NSF) grant number ACI-1053575. The research was also supported in part by NSF Grant Numbers PHY-1413799 and PHY-1806206.
\end{acknowledgments}

\bibliographystyle{apsrev4-1}

\end{document}